\documentclass[a4paper]{article}

\usepackage{INTERSPEECH2020}
\usepackage{hyperref}
\usepackage{siunitx}
\usepackage{mathtools}
\usepackage{comment}
\usepackage{enumitem}
\usepackage{threeparttable}

\newcommand{\numcircle}[1]{\textcircled{\raisebox{-.3pt}{{\scriptsize #1}}}}

\usepackage{xspace}
\makeatletter
\DeclareRobustCommand\onedot{\futurelet\@let@token\@onedot}
\def\@onedot{\ifx\@let@token.\else.\null\fi\xspace}

\def\ie{\emph{i.e}\onedot} 
 
\def\etc{\emph{etc}\onedot} 
 
\def\etal{\emph{et al}\onedot}
\makeatother

\newcommand{\abs}[1]{\left\lvert#1\right\rvert}
\newcommand{\norm}[1]{\left\lVert#1\right\rVert}
\newcommand{\herm}[1]{#1^\mathsf{H}}
\newcommand{\trans}[1]{#1^\mathsf{T}}
\newcommand{\trace}[1]{\mathrm{tr}\left\{#1\right\}}
\DeclareMathOperator*{\argmax}{arg\,max}
\def\appendixautorefname~#1\null{~#1 \null}

\title{Utterance-Wise Meeting Transcription System\\Using Asynchronous Distributed Microphones}
\name{Shota Horiguchi, Yusuke Fujita, Kenji Nagamatsu}
\address{
  Hitachi, Ltd.
}
\email{\{shota.horiguchi.wk, yusuke.fujita.su, kenji.nagamatsu.dm\}@hitachi.com}

\begin{document}
\abovedisplayskip=4pt
\belowdisplayskip=4pt
\setlength\textfloatsep{15pt}
\setlength\abovecaptionskip{5pt}
\setlength\dbltextfloatsep{11pt}

\maketitle
\begin{abstract}
  A novel framework for meeting transcription using asynchronous microphones is proposed in this paper. It consists of audio synchronization, speaker diarization, utterance-wise speech enhancement using guided source separation, automatic speech recognition, and duplication reduction. 
  Doing speaker diarization before speech enhancement enables the system to deal with overlapped speech without considering sampling frequency mismatch between microphones. Evaluation on our real meeting datasets showed that our framework achieved a character error rate (CER) of \SI{28.7}{\percent} by using 11 distributed microphones, while a monaural microphone placed on the center of the table had a CER of \SI{38.2}{\percent}.
  We also showed that our framework achieved CER of \SI{21.8}{\percent}, which is only 2.1 percentage points higher than the CER in headset microphone-based transcription.
\end{abstract}
\noindent\textbf{Index Terms}: meeting transcription, speech recognition, speaker diarization, asynchronous distributed microphones

\section{Introduction}
Meeting transcription is one practical use case of automatic speech recognition (ASR).
Difficulties are i) that input audio signals suffer from reverberation and background noise because each utterance is recorded by distant microphones and ii) that they also suffer from speech overlap because each participant speaks at any time.
To transcribe speech in such a wild condition, a powerful speech enhancement module is necessary.
Most meeting transcription systems are therefore based on a microphone array \cite{stolcke2007sri,hain2011transcribing,ito2017probabilistic,yoshioka2018recognizing}, sometimes one with an omnidirectional camera \cite{hori2011low,yoshioka2019advances} for face tracking.
This means that the system requires special equipment to be introduced.
If the microphone arrays can be replaced by more general devices, such as participants' smartphones or tablets, the usability of the system will be drastically improved.
When such devices are distributed to transcribe a meeting, the problem is that they are asynchronous, and speech separation methods for synchronized signals cannot be simply applied.

Recently, some methods of meeting transcription using asynchronous distributed microphones have been proposed.
One is the session-wise approach proposed by Araki \etal \cite{araki2017meeting,araki2018meeting}.
They first synchronized multichannel observations by solving sampling frequency mismatch, then applied session-wise speech enhancement using the minimum variance distortionless response (MVDR) beamformer, then fed the enhanced signals into an ASR module to obtain the final transcription results.
They showed that speech enhancement using asynchronous distributed microphones improved the ASR performance \cite{araki2017meeting,araki2018meeting}.
The MVDR beamformer is a frequency-wise algorithm, however, so the well-known permutation problem of frequency-domain has to be solved.
The common approach for multi-speaker cases is to prepare initial spatial correlation matrices from audio data with a fixed number of speakers and their positions \cite{higuchi2017online}.
Therefore, when the number of speakers in the inference audio is different from, especially larger than, that in the training set, we cannot provide initial spatial correlation matrices.
If we cannot obtain such spatial correlation matrices beforehand, we have to solve the permutation problem as a post-processing \cite{sawada2004robust,sawada2010underdetermined}, but there are few reports that these methods perform well on real noisy and reverberant data.

Another is the block-wise approach proposed by Yoshioka \etal \cite{yoshioka2019meeting}.
They synchronized input audio streams in a  block-online manner and then applied block-wise speech separation. The separated audios are input into the ASR module, which is followed by speaker diarization.
The benefit of this approach is that the effect of sampling frequency mismatch can be ignored within a block when the block is short enough because the scale of sampling frequency mismatch is about \SI{100}{ppm} (parts per million) at most \cite{miyabe2015blind,araki2019estimation}.
However, their speech separation uses speech-vs-noise criteria and thus cannot deal with multiple speakers speaking simultaneously.

This paper investigates the utterance-wise approach, which is different from the session-wise or block-wise approaches described above.
We first roughly synchronized audio signals recorded by distributed microphones and then applied speaker diarization.
Speaker diarization is based on the clustering of features extracted from short segments, but we use features extracted from all the signals recorded by each microphone so that it can deal with overlapped speech.
Then we applied guided source separation \cite{boeddeker2018front}, which performed well for ASR in a dinner party scenario \cite{kanda2019guided,zorila2019investigation}. This separation is conducted for each extracted utterance, which is short enough not to be suffered from sampling frequency mismatch between microphones.
We applied ASR for each enhanced utterance, and finally, we conducted duplication reduction for ASR results to reduce the effect of errors on diarization or separation.
Our approach can deal with speaker overlap without any methods to correct sampling frequency mismatch in the synchronization phase and solve the permutation problem in the speech enhancement phase.
To evaluate our framework, we recorded eight sessions of real meetings using 11 distributed smartphones, each of which was equipped with a monaural microphone.
The experimental results showed that our framework improved performance by using multiple microphones.
We also showed that our framework could achieve performance comparable to that of headset microphone-based transcription if the oracle diarization results were known.

\section{Method}
\begin{figure*}[t]
    \includegraphics[width=\linewidth]{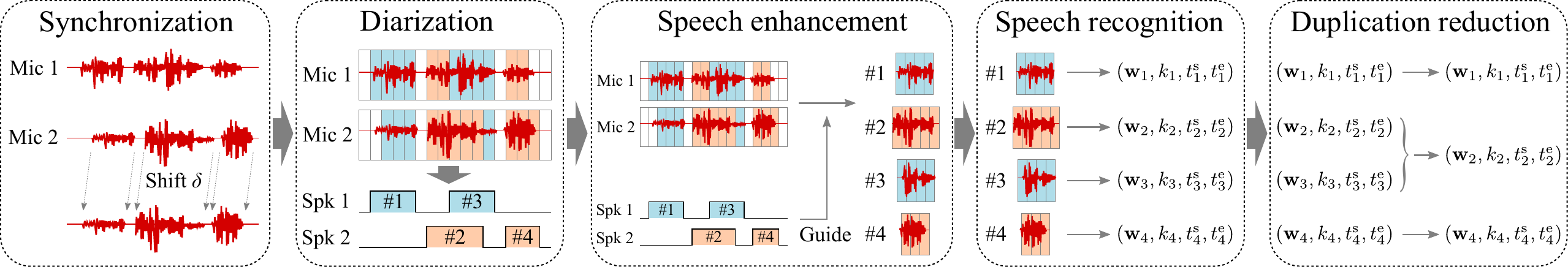}
    \caption{Overview of our meeting transcription system using asynchronous distributed microphones.}
    \label{fig:overview}
\end{figure*}

We assume that a meeting is recorded by $M$ asynchronous distributed microphones and transcription is based on the known number of speakers $K$ in an offline manner.
An overview of our method is shown in \autoref{fig:overview}. Given $M$ audio signals, we first synchronize them by maximizing their correlation. The correction of sampling frequency mismatch between signals is not conducted in the synchronization part. With the synchronized signals, we conduct clustering-based diarization to obtain utterances for each speaker. After that, we perform speech enhancement for each utterance by using the diarization results as guides to avoid the permutation problem.
The enhanced utterances are fed into the ASR module to obtain ASR results.
Finally, to reduce errors caused by diarization or separation, we apply duplication reduction for the ASR results.
In this section, we explain the details of each module of the system.

\subsection{Blind synchronization}
In this part, we conduct a correlation-based synchronization to correct start or end point differences of input signals.
This rough synchronization can be operated under the existence of the sampling frequency mismatch.
Assume that the observation of the $m$-th microphone ($m\in \{1,\dots,M\}$) is defined as $\hat{\mathbf{x}}_m\coloneqq\left[\hat{x}_{m,n}\right]_{n=1}^{N_m}$.
We select an anchor $m_a$ from the $M$ microphones and calculate the shift $\delta_m$ between signals of the anchor $m_a$ and each microphone $m\in\{1,\dots,M\}$ as follows:
\begin{gather}
    \delta_m=\begin{cases}
        \argmax_{\delta\in\mathbb{Z}}\sum_\nu x_{m_a,\nu}x_{m,\nu+\delta}&(m\neq m_a)\\
        0 & (m=m_a),
    \end{cases}\\
    x_{m,\nu}=\begin{cases}
        \hat{x}_{m,\nu}&\left(\nu\in\{1,\dots,N_m\}\right)\\
        0&\left(\mathrm{otherwise}\right).\\
    \end{cases}
\end{gather}
Synchronized signals $\mathbf{x}_m$ $(m=1,\dots,M)$ are defined in the time interval recorded by all the microphones as follows:
\begin{align}
    \mathbf{x}_m&=\left[\hat{x}_{m,n}\right]_{n=n_\text{begin}+\delta_m}^{n_\text{end}+\delta_m},\\
    n_\text{begin}&=\max_{m'\in\left\{1,\dots,M\right\}}\left(1-\delta_{m'}\right),\\
    n_\text{end}&=\min_{m'\in\left\{1,\dots,M\right\}}\left(N_{m'}-\delta_{m'}\right).
\end{align}
In this study we assume that all the utterances to be transcribed are within the time interval of $\mathbf{x}_m$.

\subsection{Speaker diarization}
In this paper, we conduct speaker diarization by clustering vectors.
One drawback of the conventional clustering-based diarization using a monaural recording is that it cannot deal with speaker overlap because each timeslot is assigned to one speaker.
On the other hand, in our scenario, each meeting has been recorded by distributed microphones.
Therefore, even when two speakers spoke simultaneously, it is expected that one microphone could have captured the one speaker's utterance at a sufficient signal-to-noise ratio (SNR) and another microphone could have captured the other speaker's utterance at a sufficient SNR.
In this study, we extract features from all the signals from all the microphones and conduct clustering for the extracted features all together to deal with speaker overlap.
 
We first split the synchronized observations $\{\mathbf{x}_m\}_m$ into short segments $\{\mathbf{x}_{m,t}\}_{m,t}$ with \SI{1.5}{\s} of window size and \SI{0.75}{\s} of window shift, where $t=1,\dots,T$ denotes the timeslot index.
We apply power-based speech activity detection for each segment; as a result, each segment is classified as either speech or non-speech.
From each speech segment, we extract features to be used for clustering.
In this study, we concatenate two kinds of features: speaker characteristics based features and power ratio based features.

For features to represent speaker characteristics, we use x-vectors \cite{snyder2018xvectors}, which are used in the state-of-the-art diarization systems \cite{sell2018diarization,diez2019bayesian}.
We extract x-vectors from the audio of each microphone so that we can obtain different speaker characteristics from the same timeslot; thus we can deal with speaker overlap. 
Before we use the vectors for clustering, we subtract a mean vector within a session from each x-vector and normalized it to have unit norm.
As a result, we obtain microphone and timeslot-wise $D$-dimensional features $\mathbf{c}_{m,t}\in\mathbb{R}^D$.

Although x-vectors from distributed microphones are potentially beneficial to diarize overlapped speech, it becomes a problem that an utterance from the same speaker could be judged as one from multiple speakers because x-vectors suffer from speaker-microphone distance and noisy environments.
Thus, we introduce power-based timeslot-wise features $\mathbf{p}_t\coloneqq\trans{\left[p_{1,t},\dots,p_{M,t}\right]}$, where $p_{m,t}$ is the average power at $\mathbf{x}_{m,t}$.
This speaker diarization part is a session-level one, so we avoid using phase-based features like GCC-PHAT \cite{knapp1976generalized} because they suffer from the sampling frequency mismatch.

Final $(D+M)$-dimensional features to be clustered are
\begin{align}
    \mathbf{v}_{m,t}=\left[
    \begin{array}{c}
        \mathbf{c}_{m,t} \\
        \lambda\mathbf{p}_{t}/\norm{\mathbf{p}_t}
    \end{array}\right],
    \label{eq:feature}
\end{align}
where $\lambda$ is the scaling factor to balance the effect of $\mathbf{c}_{m,t}$ and $\mathbf{p}_{t}$.
We apply agglomerative hierarchical clustering for the features to divide the speech segments into $K$ clusters.
As a result, each feature from a speech segment belongs to one of the clusters $\mathcal{C}_1,\dots,\mathcal{C}_K$, where $\mathcal{C}_k$ corresponds to the speech cluster of $k$-th speaker.
We also define the additional noise cluster $\mathcal{C}_{K+1}\coloneqq\{\mathbf{v}_{m,t}\}_{m,t}$.
The diarization results including noise $Y=\{y_t^{(k)}\}\in\left\{0,1\right\}^{(K+1)\times T}$ are calculated as
\begin{align}
    y_t^{(k)}=\begin{cases}
    1&\left(\exists m\in\{1,\dots,M\},~\mathbf{v}_{m,t}\in \mathcal{C}_k\right)\\
    0&\left(\mathrm{otherwise}\right).
    \end{cases}
\end{align}

In the diarization results, utterances are sometimes divided into some short fragments due to the existence of backchannels, noises, \etc.
In this study, we treat silence of \SI{1.5}{\second} or less between speech fragments from the same speaker as a speech by applying two iterations of binary closing along the time axis.

Here each timeslot in the diarization results corresponds to \SI{0.75}{\s}, which is inconsistent with the signals used in speech enhancement in the next section. 
Thus, we upsample the diarization results so that each timeslot corresponds to \SI{16}{\ms}.
Hereafter, $Y=\{y_t^{(k)}\}$ denotes the upsampled diarization results.

\subsection{Speech enhancement}
In this study, we conducted speech enhancement for each utterance by using guided source separation (GSS) \cite{boeddeker2018front}.
While the original GSS utilized oracle speech activities, we instead use estimated diarization results described in the previous section.

We first apply Weighted Prediction Error \cite{nakatani2010speech} to the input multichannel signals in a short-time Fourier transform (STFT) domain for dereverberation.
The frame length and the frame shift for the STFT were set to \SI{64}{\ms} and \SI{16}{\ms}, respectively. 
After that, speech separation by GSS \cite{boeddeker2018front} using a complex Angular Central Gaussian Mixture Model (cACGMM) \cite{ito2016complex} is applied.
Given $M$-channel observations in the STFT domain $\mathbf{X}_{t,f}\in\mathbb{C}^M$, the probability density function of the cACGMM for the signals is defined as
\begin{align}
    p\left(\hat{\mathbf{X}}_{t,f};\{\alpha_f^{(k)},B_{f}^{(k)}\}_k\right)&=\sum_k \alpha_f^{(k)}\mathcal{A}\left(\hat{\mathbf{X}}_{t,f};B_{f}^{(k)}\right),
\end{align}
where $\hat{\mathbf{X}}_{t,f}=\mathbf{X}_{t,f}/\norm{\mathbf{X}_{t,f}}$ and $\alpha_f^{(k)}$ is the mixture weight for the $k$-th source of the frequency bin $f$. $\mathcal{A}(\hat{\mathbf{X}};B)$ is a complex Angular Central Gaussian distribution \cite{kent1997data} parameterized by $B\in\mathbb{C}^{M\times M}$.
The cACGMM is optimized by the EM algorithm.
At the E-step we calculate posteriors $\gamma_{t,f}^{(k)}$ for each speaker at time-frequency bin as follows:
\begin{align}
    \gamma_{t,f}^{(k)}\leftarrow\frac{\alpha_f^{(k)}y_t^{(k)}\frac{1}{\det\left(B_f^{(k)}\right)}\frac{1}{\left[\herm{\hat{\mathbf{X}}}_{t,f}\left(B_f^{(k)}\right)^{-1}\hat{\mathbf{X}}_{t,f}\right]^M}}{\sum_{k'}\alpha_f^{(k')}y_t^{(k')}\frac{1}{\det\left(B_f^{(k')}\right)}\frac{1}{\left[\herm{\hat{\mathbf{X}}}_{t,f}\left(B_f^{(k')}\right)^{-1}\hat{\mathbf{X}}_{t,f}\right]^M}}.
    \label{eq:e_step}
\end{align}
Here the diarization result $y_t^{(k)}$ works as a guide at this E-step to force the posterior probability to be zero when the speaker $k$ does not speak at time $t$.
At the M-step the parameters $\alpha_f^{(k)}$ and $B_f^{(k)}$ are updated as follows:
\begin{align}
    \alpha_f^{(k)}\leftarrow\frac{1}{T}\sum_t\gamma_{t,f}^{(k)},\quad
    B_f^{(k)}\leftarrow M\frac{\sum_t\gamma_{t,f}^{(k)}\frac{\hat{\mathbf{X}}_{t,f}\herm{\hat{\mathbf{X}}}_{t,f}}{\herm{\hat{\mathbf{X}}}_{t,f}\left(B_f^{(k')}\right)^{-1}\hat{\mathbf{X}}_{t,f}}}{\sum_t\gamma_{t,f}^{(k)}},
\end{align}
where $\herm{(\cdot)}$ denotes Hermitian transpose.
To solve the permutation problem, \SI{15}{\s} of audios before and after each segment are used as ``context''.
After 10 iterations of optimization, we calculate the spatial covariance matrices for speech and noise as follows:
\begin{align}
    R_{f}^\text{speech}&=\frac{1}{T}\sum_t \gamma_{t,f}^{(k_\text{target})}\mathbf{X}_{t,f}\herm{\mathbf{X}}_{t,f}\in\mathbb{C}^{M\times M},\\
    R_{f}^\text{noise}&=\frac{1}{T}\sum_t \left(1-\gamma_{t,f}^{(k_\text{target})}\right)\mathbf{X}_{t,f}\herm{\mathbf{X}}_{t,f}\in\mathbb{C}^{M\times M}.
\end{align}
Here we assume that the target speaker is $k_\text{target}\in\{1,\dots,K\}$.
The MVDR beamformer $\mathbf{w}_f$ is calculated using the spatial covariance matrices as
\begin{align}
    \mathbf{w}_f&=\frac{{R_f^\text{noise}}^{-1}R_f^\text{speech}\mathbf{r}}{\trace{{R_f^\text{noise}}^{-1}R_f^\text{speech}}},\label{eq:beamformer}
\end{align}
where $\mathbf{r}$ is an one-hot vector that corresponds to the reference microphone.
Finally, Blind Analytic Normalization (BAN) postfilter \cite{warsitz2007blind} is applied for $\mathbf{w}_f$ to obtain the final beamformer, which is used for speech enhancement.
The enhanced utterance in the STFT domain is calculated as
\begin{align}
    z_{t,f}=\herm{\mathbf{w}_f}\mathbf{X}_{t,f}.
\end{align}

\subsection{Speech recognition}
For each enhanced utterance, we apply ASR consisting of a CNN-TDNN-LSTM acoustic model (AM) \cite{kanda2018lattice} followed by 4-gram-based and recurrent neural network-based language models (LMs) \cite{kanda2017investigation}.
The AM takes 40-dimensional log-scaled Mel-filterbank and 40-dimensional Mel-frequency cepstral coefficients as input audio features. 100-dimensional i-vectors are also fed into the AM for online adaptation for speaker and environment \cite{saon2013speaker}.
It was trained by 1700 hours of Japanese speech corpus using the lattice-free maximum mutual information criterion \cite{povey2016purely}.
The LMs were trained by transcriptions of the corpus used for AM training and the Wikipedia corpus.

\subsection{Duplication reduction}
The diarization and speech enhancement is not perfect, so the same transcription is sometimes included in multiple estimated utterances.
Therefore, we apply duplication reduction for the ASR results.
Widely used ensemble techniques such as ROVER \cite{fiscus1997post} and confusion network combination \cite{evermann2000posterior} are for the different ASR results obtained from the same utterance; thus, they cannot be used in this situation where the utterances to be merged have different start and end points.
To overcome this issue, we propose a combination technique for such utterances which have different time intervals.
We first find which pairs of utterances should be merged.
Given the set of $U$ ASR results $\mathcal{W}=\{(\mathbf{w}_u, k_u, t_u^\mathrm{s}, t_u^\mathrm{e})\}_{u=1}^U$, where $\mathbf{w}_u$, $k_u$, $t_u^\mathrm{s}$, and $t_u^\mathrm{e}$ denote the sequence of words, speaker, start time, and end time of $u$-th result, respectively, we calculate an adjacency matrix $A=\{a_{i,j}\}_{i,j}\in\{0,1\}^{U\times U}$ as follows:
\begin{align}
    a_{i,j}&=\begin{cases}
        1&(\max{(t_i^{\mathrm{s}},t_j^{\mathrm{s}})}<\min{(t_i^{\mathrm{e}}, t_j^{\mathrm{e}})}~\land\\
        &\quad s\left(\mathbf{w}_i, \mathbf{w}_j\right)>\tau~\land~k_i\neq k_j)\\
        0&(\mathrm{otherwise}),
    \end{cases}
    \label{eq:adjacency_matrix}
\end{align}
where $\tau\in[0,1]$ is the threshold value.
Here $s\left(\mathbf{w}_i,\mathbf{w}_j\right)$ is the similarity between $\mathbf{w}_i$ and $\mathbf{w}_j$ defined as follows:
\begin{align}
    s\left(\mathbf{w}_i,\mathbf{w}_j\right)\coloneqq\frac{\max\left(\abs{\mathbf{w}_i},\abs{\mathbf{w}_j}\right)-d\left(\mathbf{w}_i,\mathbf{w}_j\right)}{\min\left(\abs{\mathbf{w}_i},\abs{\mathbf{w}_j}\right)},
\end{align}
where $d(\mathbf{w}_i, \mathbf{w}_j)$ is the Levenshtein distance between $\mathbf{w}_i$ and $\mathbf{w}_j$, and $\abs{\mathbf{w}}$ denotes the number of words in $\mathbf{w}$.
With this adjacency matrix, all the elements in $\mathcal{W}$ can be clustered into $C$ clusters. We denote the clustering result as $\mathcal{C}=\{c_u\}_{u=1}^{U}\in\{1,\dots,C\}^U$, which fulfill $c_i=c_j$ if a path between $i$-th and $j$-th elements exists in $A$ and $c_i\neq c_j$ otherwise.
Assuming that $\mathcal{W}_{k,c}\subseteq\mathcal{W}$ is the set of ASR results which belong to the cluster $c$ and are uttered by speaker $k$, we obtain the representative speaker $k^c$ of the cluster $c$ by
\begin{align}
    k^c&=\argmax_{k\in\{1,\dots,K\}}f(\mathcal{W}_{k,c}),
\end{align}
where $f(\cdot)$ is the selection function. 
In this study, we select the speaker with the longest utterance(s), \ie, $f(\mathcal{W}_{k,c})=\sum_{(\mathbf{w},k,t^\mathrm{s},t^\mathrm{e})\in\mathcal{W}_{k,c}}{\abs{\mathbf{w}}}$.
The set of de-duplicated ASR results $\mathcal{W}'$ can be obtained as follows:
\begin{align}
    \mathcal{W}'&=\bigcup_{c\in\{1,\dots,C\}}\mathcal{W}_{k^c,c}.
\end{align}

\section{Experiments}
\subsection{Data}
\begin{figure}[t]
    \centering
    \includegraphics[width=0.8\linewidth]{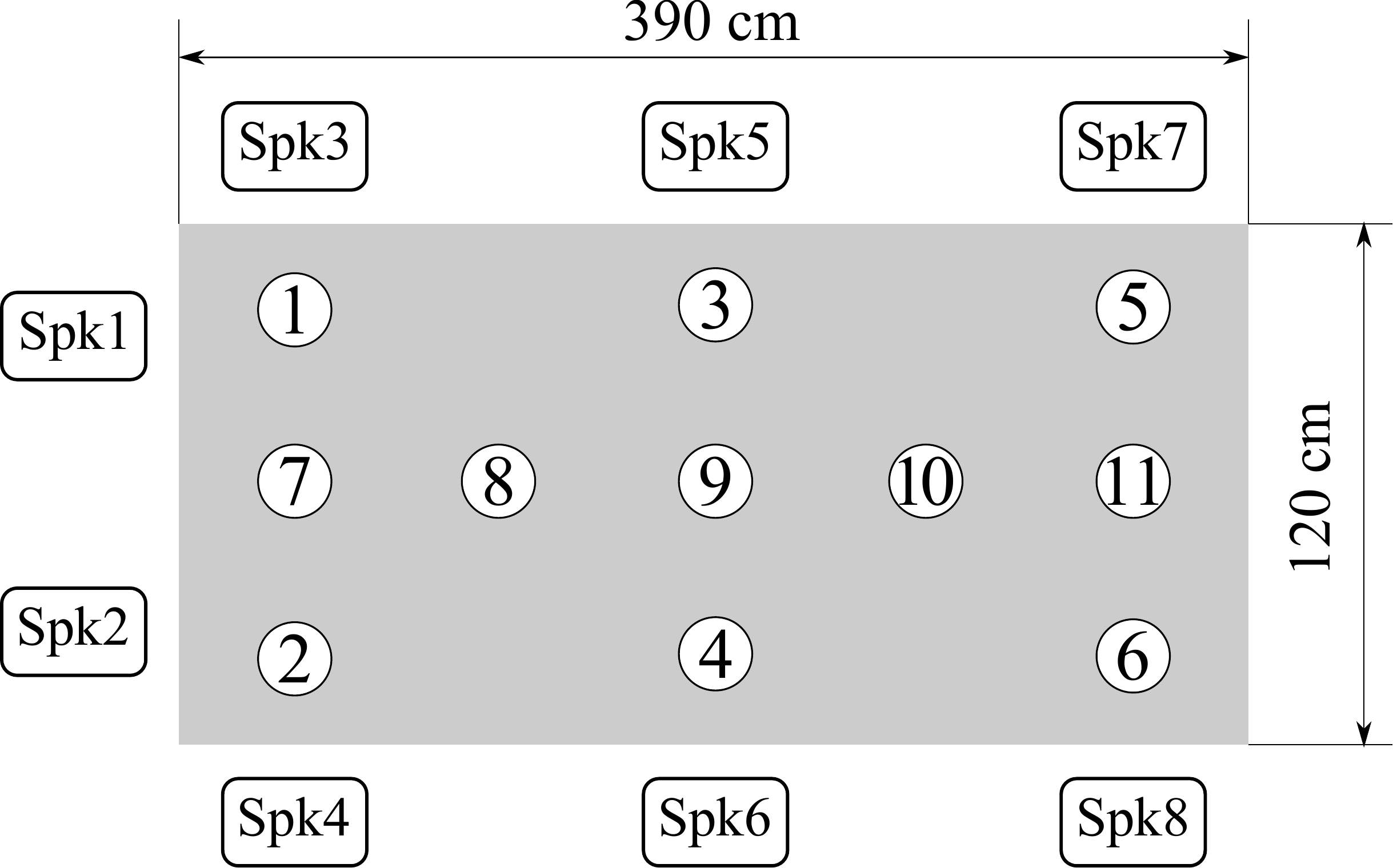}
    \caption{Recording environment. \numcircle{1}-\numcircle{11} denote smartphones, each of which is equipped with a monaural microphone.}
    \label{fig:recording_environment}
\end{figure}
To evaluate the performance of our method, we collected eight sessions of real meeting data.
The recording environment is shown in \autoref{fig:recording_environment}.
Each session had at most eight participants and was recorded by 11 smartphones distributed on the table. Each smartphone was equipped with a monaural microphone to record meetings at \SI{16}{\kHz} / 16 bit.
Each participant wore a headset microphone, and the groundtruth transcriptions were based on the headset recordings.
The statistics of collected data are shown in \autoref{tbl:datasets}.
The recordings correspond to about two hours of meetings with an average overlap ratio of \SI{13.2}{\percent}.

\begin{table}[t]
    \centering
    \caption{Statistics of the recorded meetings.}
    \label{tbl:datasets}
    \resizebox{\linewidth}{!}{
    \begin{tabular}{@{}cccS[table-format=3]S[table-format=2.1]@{}}
        \toprule
        Session ID & \#Speakers & {Duration} & {\#Utterances} & {Overlap ratio (\%)}\\\midrule
        I&7&19:49&160&6.9\\
        I\hspace{-.1em}I&8&14:27&150&14.0\\
        I\hspace{-.1em}I\hspace{-.1em}I&5&13:13&198&16.6\\
        I\hspace{-.1em}V&7&12:08&184&19.9\\
        V&6&12:37&80&5.5\\
        V\hspace{-.1em}I&6&16:50&256&14.7\\
        V\hspace{-.1em}I\hspace{-.1em}I&7&16:25&223&11.3\\
        V\hspace{-.1em}I\hspace{-.1em}I\hspace{-.1em}I&7&11:25&185&19.9\\\midrule
        Avg.&-&116:54&1436&13.2\\
        \bottomrule
    \end{tabular}
    }
\end{table}

\subsection{Results}
We investigated various combinations of asynchronous distributed microphones: 2 microphones (\numcircle{8}\&\numcircle{10} in \autoref{fig:recording_environment}), 3 microphones (\numcircle{7}\&\numcircle{9}\&\numcircle{11}), 6 microphones (\numcircle{1}-\numcircle{6}), and 11 microphones (\numcircle{1}-\numcircle{11}).
For comparison, we also evaluated the performance of one monaural microphone (\numcircle{9}) and of headset microphones that the participants wore during each session.

The character error rates (CERs) obtained using various microphone combinations in each session are shown in \autoref{tbl:asr_results}.
In these experiments, the weighting parameter $\lambda$ in \autoref{eq:feature} was set to $1.0$.
By using multiple microphones, we could have reduced CERs, especially by using a large number of microphones.
Note that in two-, three-, and six-microphone settings, using more microphones not always resulted in better CERs.
This is because the sets of microphones in these settings are disjoint and the CERs highly depended on the positions of microphones and speakers.
On the other hand, we observed the best CERs in almost every session by using all the 11 microphones. 
This result indicated that adding microphones has almost no negative effect on CERs.
In \autoref{tbl:asr_results}, we also showed CERs with 11 microphones in the case when oracle diarization was used for GSS.
It achieved the CER of \SI{21.8}{\percent}, which is only 2.1 percentage points worse than the CER of \SI{19.7}{\percent} obtained using headset microphones.
It can be said that our method can potentially achieve nearly headset-level CERs when it is used with a more powerful diarization method \cite{fujita2019end2,medennikov2020stc,horiguchi2020endtoend}.

In \autoref{tbl:various_weights} we show the average CERs over sessions with various weighting parameters $\lambda$ in \autoref{eq:feature}.
Combinations of speaker characteristics based features and power ratio based features improved transcription performance, especially when the number of microphones is smaller and the power ratio thus has less information about the directions of speakers.

Finally, we conducted ablation studies by removing binary closing in diarization, speech enhancement by using recordings of the reference microphone instead, and duplication reduction, respectively. 
Here we used 11 microphones with $\lambda=1.0$.
The results are shown in \autoref{tbl:ablation_study}.
We found 1.9, 9.1, and 3.2 percentage points degradation from the baseline by removing binary closing, speech enhancement, and duplication reduction, respectively.
From there results, we concluded that these three components contributed to the improvement of the CER.

\begin{table}[t]
    \begin{minipage}[t]{\linewidth}
        \centering
        \caption{CERs (\%) obtained using various microphone combinations.}
        \label{tbl:asr_results}
        \begin{threeparttable}
        \resizebox{\linewidth}{!}{
        \begin{tabular}{@{}rccccccccc@{}}
            \toprule
            &\multicolumn{8}{c}{Session ID}\\\cmidrule(lr){2-9}
            {\#Mic}&I&I\hspace{-.1em}I&I\hspace{-.1em}I\hspace{-.1em}I&I\hspace{-.1em}V&V&V\hspace{-.1em}I&V\hspace{-.1em}I\hspace{-.1em}I&V\hspace{-.1em}I\hspace{-.1em}I\hspace{-.1em}I&Avg.\\\midrule
            1&31.2&30.1&37.1&37.6&28.2&48.4&50.4&52.5&38.2\\
            2 &22.9&25.3&30.5&37.0&21.8&41.7&36.8&45.7&31.4\\
            3 &26.8&24.4&35.9&37.2&23.2&43.1&41.9&46.6&33.7\\
            6 &22.3&22.2&36.0&32.1&21.0&38.1&35.1&44.3&30.2\\
            11&21.2&21.1&32.5&30.9&19.6&37.6&34.0&41.0&28.7\\
            11\tnote{*}&17.0&16.3&21.7&21.2&17.7&27.0&27.0&32.8&21.8\\\midrule
            {Headset}&18.3&15.8&21.0&20.1&13.6&21.3&24.9&25.8&19.7\\
            \bottomrule
        \end{tabular}
        }
        \begin{tablenotes}
            \item[*] The oracle diarization was used for speech enhancement.
        \end{tablenotes}
        \end{threeparttable}
        \vspace{1em}
    \end{minipage}\\
    \begin{minipage}[t]{\linewidth}
        \centering
        \caption{CERs (\%) obtained with various scaling factors $\lambda$ in \autoref{eq:feature}.}
        \label{tbl:various_weights}
        \scalebox{0.9}{
        \begin{tabular}{@{}rccccccc@{}}
            \toprule
            &\multicolumn{7}{c}{$\lambda$}\\\cmidrule(l){2-8}
            \#Mic&$2^{-3}$&$2^{-2}$&$2^{-1}$&$2^0$&$2^{1}$&$2^{2}$&$2^{3}$\\\midrule
            2&33.7&31.7&32.2&\textbf{31.4}&31.8&33.3&35.5\\
            3&34.2&34.3&34.0&33.7&\textbf{33.0}&34.8&35.2\\
            6&33.5&34.1&33.4&\textbf{30.2}&31.4&31.9&32.4\\
            11&33.5&32.5&31.1&28.7&28.9&\textbf{28.4}&28.9\\
            \bottomrule
        \end{tabular}
        }
        \vspace{1em}
    \end{minipage}
    \begin{minipage}[t]{\linewidth}
        \centering
        \caption{Ablation study using 11 microphones.}
        \label{tbl:ablation_study}
        \scalebox{0.9}{
        \begin{tabular}{@{}lc@{}}
            \toprule
            Method&CER (\%)\\\midrule
            Baseline (11 mics)& 28.7\\
            w/o binary closing&30.6\\
            w/o speech enhancement &37.8\\
            w/o duplication reduction &31.9\\
            \bottomrule
        \end{tabular}
        }
    \end{minipage}
\end{table}

\section{Conclusions}
In this paper, we proposed a meeting transcription system based on utterance-wise processing using asynchronous distributed microphones.
It consists of the following modules: blind synchronization, speaker diarization, speech enhancement, speech recognition, and duplication reduction.
Evaluation on the real meeting data showed the effectiveness of our framework and its components, and also showed that it could perform comparably to the headset microphone-based transcription if the oracle diarization was given.
The future perspective of this research is to operate this framework in an online manner.

\bibliographystyle{IEEEtran}
\bibliography{mybib}

\end{document}